\documentclass[twocolumn]{aastex631}

\usepackage{amsmath}
\usepackage{graphicx}
\usepackage{textcomp}
\usepackage{gensymb}
\usepackage{natbib}
\usepackage{longtable}
\usepackage{multirow}
\usepackage{float}
\usepackage{enumitem}

\begin{document}
\def\teff{$T_{\rm eff}$}
\def\cs{$\chi^{2}$}
\def\rsun{$R_{\odot}$}
\def\msun{$M_{\odot}$}
\def\rstar{$R_{\star}$}
\def\rearth{$R_{\earth}$}
\def\av{$A_{V}$}
\def\emcee{\texttt{emcee}}
\def\kepler{\textit{Kepler}}
\def\tpri{$T_{\rm eff 1}$}
\def\tsec{$T_{\rm eff 2}$}
\def\rpri{$R_{\star 1}$}
\def\rsec{$R_{\star 2}$}

\graphicspath{{figures/}}
\shorttitle{Binary Planet Hosts IV.}
\shortauthors{Sullivan et al.}

\title{Revising Properties of Planet-Host Binary Systems. IV. The Radius Distribution of Small Planets in Binary Star Systems is Dependent on Stellar Separation
\footnote{Based on observations obtained with the Hobby-Eberly Telescope, which is a joint project of the University of Texas at Austin, the Pennsylvania State University, Ludwig-Maximilians-Universität München, and Georg-August-Universität Göttingen.}}

\author[0000-0001-6873-8501]{Kendall Sullivan}
\affil{Department of Astronomy and Astrophysics, University of California Santa Cruz, Santa Cruz, CA 95064, USA}

\author[0000-0001-9811-568X]{Adam L. Kraus}
\affil{Department of Astronomy, University of Texas at Austin, Austin, TX 78712, USA}


\author[0000-0002-2580-3614]{Travis A. Berger}
\affiliation{Space Telescope Science Institute, 3700 San Martin Drive, Baltimore, MD 21218, USA}

\author[0000-0001-9823-1445]{Trent J. Dupuy}
\affiliation{Institute for Astronomy, University of Edinburgh, Royal Observatory, Blackford Hill, Edinburgh, EH9 3HJ, UK}

\author[0009-0009-6539-2432]{Elise Evans}
\affiliation{Institute for Astronomy, University of Edinburgh, Royal Observatory, Blackford Hill, Edinburgh, EH9 3HJ, UK}

\author[0000-0002-5258-6846]{Eric Gaidos}
\affiliation{Department of Earth Sciences, University of Hawai'i at M\"{a}noa, Honolulu, HI 96822, USA}

\author[0000-0001-8832-4488]{Daniel Huber}
\affiliation{Institute for Astronomy, University of Hawai`i, 2680 Woodlawn Drive, Honolulu, HI 96822, USA}
\affiliation{Sydney Institute for Astronomy (SIfA), School of Physics, University of Sydney, NSW 2006, Australia}

\author[0000-0002-6194-043X]{Michael J. Ireland}
\affiliation{Research School of Astronomy and Astrophysics, Australian National University, Canberra ACT 2611, Australia}

\author[0000-0003-3654-1602]{Andrew W. Mann} %
\affiliation{Department of Physics and Astronomy, The University of North Carolina at Chapel Hill, Chapel Hill, NC 27599, USA}

\author[0000-0003-0967-2893]{Erik A. Petigura}
\affiliation{Department of Physics \& Astronomy, University of California Los Angeles, Los Angeles, CA 90095, USA}

\author[0000-0001-5729-6576]{Pa Chia Thao}
\affiliation{Department of Physics and Astronomy, The University of North Carolina at Chapel Hill, Chapel Hill, NC 27599, USA} 

\author[0000-0001-7336-7725]{Mackenna L. Wood}%
\affiliation{MIT Kavli Institute for Astrophysics and Space Research Massachusetts Institute of Technology, Cambridge, MA 02139, USA}

\author[0000-0002-2696-2406]{Jingwen Zhang}
\altaffiliation{NASA FINESST Fellow}
\affiliation{Institute for Astronomy, University of Hawai`i, 2680 Woodlawn Drive, Honolulu, HI 96822, USA}

\correspondingauthor{Kendall Sullivan}
\email{ksulliv4@ucsc.edu}

\begin{abstract}
Small planets ($R_{p} \leq 4$ \rearth) are divided into rocky super-Earths and gaseous sub-Neptunes separated by  a radius gap, but the mechanisms that produce these distinct planet populations remain unclear. Binary stars are the only main-sequence systems with an observable record of the protoplanetary disk lifetime and mass reservoir, and the demographics of planets in binaries may provide insights into planet formation and evolution. To investigate the radius distribution of planets in binary star systems, we observed 207 binary systems hosting 283 confirmed and candidate transiting planets detected by the Kepler mission, then recharacterized the planets while accounting for the observational biases introduced by the secondary star. We found that the population of planets in close binaries ($\rho \leq 100$ au) is significantly different from the planet population in wider binaries ($\rho > 300$ au) or single stars. In contrast to planets around single stars, planets in close binaries appear to have a unimodal radius distribution with a peak near the expected super-Earth peak of $R_{p} \sim$ 1.3 \rearth\ and a suppressed population of sub-Neptunes. We conclude that we are observing the direct impact of a reduced disk lifetime, smaller mass reservoir, and possible altered distribution of solids reducing the sub-Neptune formation efficiency. Our results demonstrate the power of binary stars as a laboratory for exploring planet formation and as a controlled experiment of the impact of varied initial conditions on mature planet populations.
\end{abstract}

\section{Introduction}
The number of known exoplanets is large and increasing steadily. With this growing population of planets, it has become possible to not only characterize individual systems but also explore the properties of planet populations. The Kepler mission \citep{Borucki2010} was designed to facilitate demographic studies, and produced a sample of planets that remains the gold standard for exoplanet demographics \citep[e.g.,][]{Winn2018}.

The properties of planet populations revealed by exoplanet demographics inform our understanding of planet formation and evolution, since any theory of planet formation must successfully predict the demographics we observe around main sequence stars. For example, demographic studies have revealed that large, close-in, gas giant planets are intrinsically rare, while smaller planets in the super-Earth to sub-Neptune regime ($1.2 R_{\oplus} \leq R_{p} \leq 4 R_{\oplus}$) are common \citep[see various occurrence rates in the literature for both large and small planets; e.g.,][]{Batalha2013, Foreman-Mackey2014, Burke2015, Mulders2015, Kunimoto2020, Dattilo2023}; that highly-irradiated planets mostly lose their atmospheres \citep[the Neptune desert;][]{Mazeh2016}; and that small rocky planet cores rarely retain substantial atmospheres \citep[the radius gap;][]{Fulton2017, VanEylen2018, Hardegree-Ullman2020, Petigura2022, Rogers2023}. Each of these observations of mature planet populations has informed our understanding of early planet formation and evolution.

Although many features of the mature exoplanet population are known, it remains challenging to connect observations to planet formation theory. One difficulty is that dynamical evolution can cause planets to migrate, be ejected, or be destroyed, which changes the system architecture \citep[e.g.,][]{Lissauer1987, Rasio1996, Chambers1996, Lin1997}. Another challenge is that it is difficult to connect the initial conditions of planet formation (the lifetime, mass, and radius of the protoplanetary disk) to the resulting planetary system, because there is no record of the disk properties in the system after the disk dissipates. Young planets provide an opportunity to explore planet formation in situ \citep[e.g.,][]{David2016, Donati2016, Mann2016, David2019, Tofflemire2021, Wood2023}, but the known population of young planets is small, meaning that demographic studies are not yet possible. Resolving the difficulty of linking initial conditions to mature populations will reveal the impact of altered conditions on mature planets and allow calibration of planet formation theories.

One possible inroad to establishing connections between protoplanetary disk properties and the resulting exoplanet population is to study planets in binary star systems (S-type or circumstellar planets in binaries; bodies that orbit one star in the system with an exterior stellar companion). Binary stars are common: about 50\% of Sun-like stars are binary \citep{Duquennoy1991, Raghavan2010, Moe2017}. Binaries impact the planet-forming environment by dynamically disrupting the disk, which outwardly truncates it \citep{Artymowicz1994, Harris2012}, potentially reducing the mass of icy solids in the outer disk; reduces the disk lifetime by about a factor of 10 for a majority of systems \citep{Cieza2009, Kraus2012}; and may reduce the disk dust mass \citep{Zurlo2020}. Two-thirds of close binary systems have lost their disk by ages of $\tau \sim 1$ Myr \citep{Kraus2012}. Multiplicity may also suppress planet formation, with small-separation binaries ($\rho \lesssim$ 100 au) possibly hosting about two-thirds fewer planets than wider systems \citep{Wang2015, Kraus2016, Moe2021}. 

The population of planets that do form and survive in close binary systems may differ from the single star population because of the altered conditions within the shorter-lived, lower-mass disk. The disk properties are driven by the binary star separation, with the disks of the closest binaries experiencing the largest impact from multiplicity \citep{Kraus2012}, so the maximum radial extent of the disk and maximum probable disk lifetime can be inferred from the properties of the mature binary system, even billions of years after the disk has dissipated. Thus, planets in main-sequence binaries provide a direct link between disk initial conditions and mature planet populations that are relatively easy to observe.

A planet population of particular interest to formation models is small planets (planets with $R_{p} \leq 4$ \rearth). Small planets are typically divided into (super-)Earths, thought to be primarily rocky with little to no atmosphere, and (sub-)Neptunes, thought to be a combination of rocky cores with gaseous hydrogen/helium atmospheres and ``water worlds'' that have a substantial water mass fraction. Super-Earths and sub-Neptunes are separated by a feature known as the radius gap \citep[or radius valley; e.g.,][]{Fulton2017, Petigura2022, Ho2023}, which is a gap in the planet radius distribution around 1.8 \rearth\ that is a result of the large radius difference between a bare or stripped rocky planet and one with even a small H/He atmosphere ($\gtrsim 1\%$ by mass). 

Most planets are thought to form with primordial H/He atmospheres, meaning that atmospheric mass loss is often invoked to explain the presence of the radius gap. The cause of the radius gap remains unclear, with a variety of suggested mechanisms that likely all contribute in some amount \citep{Owen2023}: ``boil-off'' \citep[initial atmospheric mass loss as the disk dissipates; e.g.,][]{Owen2016}; photoevaporation \citep[e.g.,][]{Owen2013}; core-powered mass loss \citep[caused by internal heating as the planetary core contracts and cools; e.g.,][]{Owen2016, Ginzburg2018, Gupta2019}; impact-driven mass loss \citep[e.g.,][]{Biersteker2019}; and late-time gas-poor planet formation \citep[e.g.,][]{Lee2022} or primordially bare planets that never accumulated atmospheres \citep[e.g.,][]{Lopez2018}. Binaries may provide a measure of the timescale and mass reservoir that produced a specific final outcome of planet formation, meaning that changes to the planet populations in different binary separation regimes may reveal the dominant mechanism that produces super-Earths and sub-Neptunes.

In previous work \citep{Sullivan2023}, we performed an initial exploration of the radius distribution of small planets in binary systems and did not detect a statistically significant radius gap in a sample of 120 planets with radii $R_{p} \leq 2.5 R_{\oplus}$. We suggested that this could indicate the planet core mass distribution changing as a function of the binary separation because of the reduced timescale for planet formation in binaries, causing the nominal single-star radius gap to appear filled in. Our sample was not large enough to explore this hypothesis by separating wide and close binaries, but we predicted that the two binary separation regimes would have different planetary radius distributions. Specifically, we suggested that the locations of the radius gap and the super-Earth peak would both shift to smaller radii with closer binary separations.

In this work, we have expanded the sample of our previous study by more than a factor of two to include 283 planets in 207 binary star systems with projected separations ranging from $\sim 10-1000$ au and planetary radii ranging from $0.5 - 4$ \rearth. We observed the new systems using the LRS2 instrument on the Hobby-Eberly Telescope (HET) at McDonald Observatory, then revised the stellar and planetary properties to account for the observational biases introduced by the presence of a secondary star using the methods of \citet{Sullivan2022b}. We used the expanded sample to investigate the planet radius distribution as a function of binary separation and connect the impact of reduced timescales and mass reservoirs for planet formation to observed exoplanet demographics.

\section{Sample Assembly and Observations}
\subsection{Sample Selection and Archival Data}\label{sec:sample}
\begin{deluxetable}{ccc}
\tablecaption{Excluded Systems \label{tab:excluded}}
\setcounter{table}{0}
\tablecolumns{3}
\tablehead{\colhead{KOI} & \colhead{Reason for exclusion} & \colhead{Source}}
\startdata
KOI-0005 &	Triple & 	Visual inspection \\
KOI-0126 &	log(g) $<$ 4 & 	Gaia \\
KOI-0200 &	Single & 	Visual inspection \\
KOI-0353 &	Triple & 	Visual inspection \\
KOI-0379 &	Triple & 	Visual inspection \\
\enddata
\tablecomments{The full table is available in the supplementary material.}
\end{deluxetable}

To select the sample we began with the list of Kepler Objects of Interest (KOIs) from the NASA Exoplanet Archive, in which the majority of sources are from Kepler DR25 \citep{Thompson2018}, who assembled a uniform sample of KOIs after the final Kepler quarter of data was released. We removed the flagged false positives (FPs) from the list, leaving a sample with 60\% Confirmed Planets and 40\% Planet Candidates. We cross-matched the KOIs with a list of binary stars (see below) with projected separations $\rho \leq 2$\arcsec and a contrast $\Delta m_{\lambda} \leq 3.5$ mag between the primary and secondary star in at least one filter. We chose these cuts to include only systems where the presence of a secondary star would be most likely to produce a bias in the stellar characterization given the 4'' pixel size of Kepler, and where high-contrast imaging is likely to be highly complete. We excluded any known triples or higher-order multiple systems with a companion within 4\arcsec\ {to ensure minimal contamination from the companion in the Kepler light curve}.

We vetted publicly available imaging from our own observing program or the data presented in \citet{Furlan2017} using the Keck Observatory Archive or the Exoplanet Follow-up Program (ExoFOP) by eye to check for any deviations from a binary system, including possible singles and higher-order multiples. Later in the analysis we also excluded any planets with radii from \citet{Thompson2018} with more than 25\% error to ensure that the transit fits were high quality. We removed any apparent binaries that were resolved in Gaia DR3 but not comoving. We removed systems that had a Gaia DR3 \citep{Gaia2023} surface gravity $\log(g) < 4$. All potentially eligible systems that we excluded are listed in Table \ref{tab:excluded} along with the reason for exclusion.

We assembled our list of binary stars from a variety of sources. We primarily used the catalog of binary stars from \citet{Furlan2017}, which in turn was a compilation of several sources of high-resolution imaging of KOIs. We supplemented this catalog with Robo-AO imaging presented in \citet{Ziegler2018} and new adaptive optics (AO) imaging using NIRC2 at Keck Observatory. Each of these methods has different sensitivity: speckle imaging is sensitive to relatively close-in brighter companions, and AO imaging is sensitive to small-separation, redder, fainter companions. Because our targets were mostly comprised of bright secondary stars we expect that all techniques should be sensitive to the majority of companions.

\begin{figure}
    \plotone{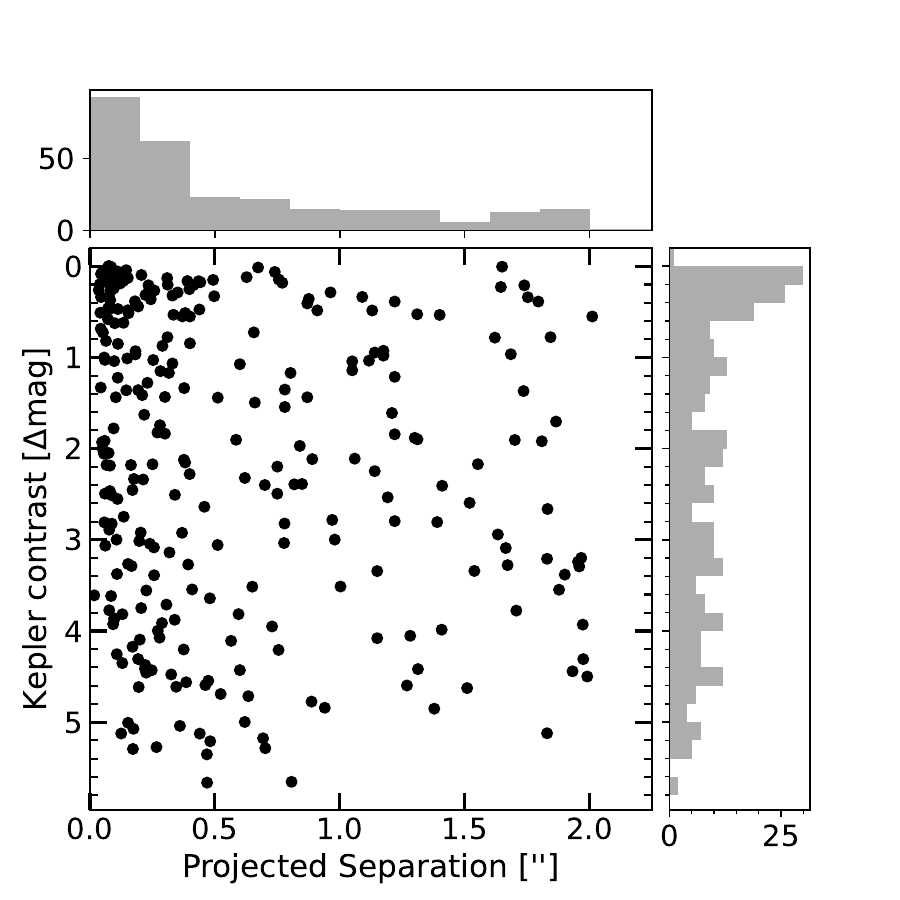}
    \caption{The angular separation and estimated contrast in the Kepler bandpass for all systems in our sample. The systems are distributed through contrast-separation space, with a high concentration of bright, small-separation secondaries.}
    \label{fig:rho_deltamag}
\end{figure}

Table \ref{tab:obs} summarizes some details of our observations as well as the accumulated contrast measurements from the literature that we used in our analysis. The table includes the KOI number of each object, the angular separation, the $r'$ magnitude from the KIC, the S/N and date of the spectroscopic observation, the contrast measurements used in the analysis, and the literature references as determined by \citet{Furlan2017} if relevant. The table includes data from \citet{Furlan2017}, which was assembled from new observations in that paper as well as observations from \citet{Adams2012}; \citet{Howell2011}; \citet{Baranec2016}; \citet{Lillo-Box2012}; \citet{Lillo-Box2014}; and \citet{Kraus2016}. Figure \ref{fig:rho_deltamag} shows the distribution of our sample in angular separation-contrast space, after calculating the approximate contrast in the Kepler bandpass.

\subsection{New High-Spatial-Resolution Imaging with Keck/NIRC2} 
We present new AO imaging of KOIs using the NIRC2 adaptive optics camera on the Keck-II telescope, obtained as part of an ongoing survey of the multiplicity of Kepler planet-hosting stars. The typical observing strategy was described by \citet{Kraus2016}. To briefly recap, a modest number of 20 sec exposures (typically 4--8, with more in marginal conditions) were obtained using a mix of natural guide star and laser guide star adaptive optics on nights spanning 2015--2022, using the narrow camera and the $K'$ broadband filter centered at 2.2 $\mu$m.

The images were analyzed as also described by \citet{Kraus2016}. To briefly summarize, the images were dark-subtracted with mode-matched darks that used identical numbers of Fowler samples and on-chip coadds, then linearized, flatfielded, and screened for cosmic rays and known hot/dead pixels.
The relative astrometry and photometry of each binary was fit within each image by first testing the $\chi^2$ goodness-of-fit for a library of potential empirical templates (encompassing the 1000 images of single stars that were taken in the same filter and closest to that date) using an initial estimate of the photometry/astrometry, then using the best empirical template to optimize the fit of the photometry and astrometry via a modified Metropolis-Hastings fitter where uphill jumps were disabled. The fit process iterated between these two stages until the same empirical template produced the same best-fit values. The values for the individual frames were then corrected for geometric distortion using the distortion solution of \citet{Service2016}, and finally averaged to produce the final astrometry and photometry, with the RMS being adopted as the uncertainty, adding the systematic uncertainty in the distortion solution (1.4 mas) in quadrature.
The new astrometry and relative contrasts are listed in Table \ref{tab:new NIRC2}. 


\subsection{Spectroscopic Observations with HET/LRS2}
We took spectroscopic observations of the full sample of KOIs with the Low-Resolution Spectrograph (LRS) on the 10 meter Hobby-Eberly Telescope (HET; \citealt{Ramsey1998, Hill2021}) between 2021 April 02 and 2023 September 26 at McDonald Observatory. The observations were taken in queue mode, and we observed all eligible targets over the course of the six trimesters (spread over 3 years) that we obtained data. We observed our sample with the red/far-red arm of the second-generation Low-Resolution Spectrograph (LRS2-R; \citealt{Chonis2014, Chonis2016}), which has a wavelength range of 6500 $\leq \lambda \leq $ 10500 \AA\ and R $\sim$ 1800. Because of severe telluric contamination and low S/N in the far-red arm, we only used the red arm for our analysis (6500 $\leq \lambda \leq $ 8470 \AA). 

LRS2 is an integral field spectrograph (IFS) that is continuously tiled with 0\farcs6 hexagonal lenslets, meaning that LRS2 observes an image with complete spatial coverage (100\% filling factor; no gaps between fibers or lenslets) at each wavelength. This feature of the instrument makes data extraction straightforward. Although we were observing binaries, we were not trying to resolve the binary components, so we set high limits on the seeing of 2\farcs5, and allowed observations during bright time and with spectroscopic transparency (50\% or higher was required). Because the systems were unresolved, we used the standard facility pipeline to reduce the data and extract the spectra. The pipeline identifies the source using the wavelength slice with the highest S/N, then extracts it using an aperture with a diameter of 2.5 times the seeing (G. Ziemann, private communication). The dates of our observations and the S/N of the spectra are summarized in Table \ref{tab:obs}.

Following observations we had to correct for atmospheric contamination in the spectra. As discussed in \citet{Sullivan2022c} and \citet{Sullivan2023}, we did not observe telluric standard stars during our observations, but instead chose to correct for atmospheric contamination after the fact using atmospheric models. This choice was motivated by the observation that the HET mirror is at a fixed altitude, so all observations are taken at a constant airmass. This means that the largest impact on the degree of telluric contamination is fluctuation of the atmospheric water content, and thus only one variable is needed to fit a telluric model to the data. 

We generated a grid of models of atmospheric contamination using \texttt{telfit} \citep{Gullikson2014} with the appropriate spectroscopic resolution and varying amounts of humidity. We performed a dual-component fit to the data using a composite spectrum created from an initial guess humidity and the observed stellar temperature from the ExoFOP. We performed a reduced-$\chi^{2}$ fit to find the best-fit humidity and then divided out the telluric model from each spectrum. We masked regions where the telluric correction was poor or had large residuals, as described in \citet{Sullivan2022c}.

\section{Star and Planet Parameter Revision}
Using our new and archival data, we revised the stellar parameters by fitting the assembled data set using a two-component spectral model, then used the new stellar parameters to revise the planetary parameters.

\subsection{Stellar Parameter Retrieval}
We analyzed the spectra and photometry using the method presented in \citet{Sullivan2022b} and updated in \citet{Sullivan2023}. We summarize the analysis here for completeness. We assembled a data set consisting of photometry from the Kepler Input Catalog \citep[KIC;][]{Brown2011} and spectroscopy from LRS2/HET (this work and \citealt{Sullivan2023}), both of which were composite measurements of the unresolved binary; and resolved flux ratios (or contrasts) from high-contrast optical speckle (literature), NIR AO imaging (literature or this work), and resolved photometry from Gaia DR3 \citep{Gaia2023} when available. We fit the data using a two-component stellar model that included extinction and distance terms to retrieve the following parameters: $\theta = $\{\tpri, \tsec, \rpri, \rsec, \av, $\varpi$\}, where \tpri\ and \tsec\ are the primary and secondary star temperatures, \rpri\ and \rsec\ are the radii, \av\ is the V-band extinction, and $\varpi$ is the parallax.

To perform the fits we generated a comparison model with synthetic photometry using the \texttt{pyphot} package \citep{pyphot}, synthetic contrasts, and a composite system spectrum matching the LRS2 data in wavelength and resolution. To generate the synthetic data we used the BT-Settl models \citep{Allard2013, Rajpurohit2013, Allard2014, Baraffe2015} with the \citet{Caffau2011} line list. We fit the data set in two stages: first, we ran an optimization with an ensemble of walkers and a modified Gibbs sampler, then we retrieved the parameters using \texttt{emcee} \citep{Foreman-Mackey2013}. We initialized the \texttt{emcee} run with the top third of the walkers from the optimization stage (the third of walkers with the lowest reduced $\chi^{2}$), which was 50 \texttt{emcee} walkers selected from the 150 walkers in the optimization step. 

\begin{figure}
    \plotone{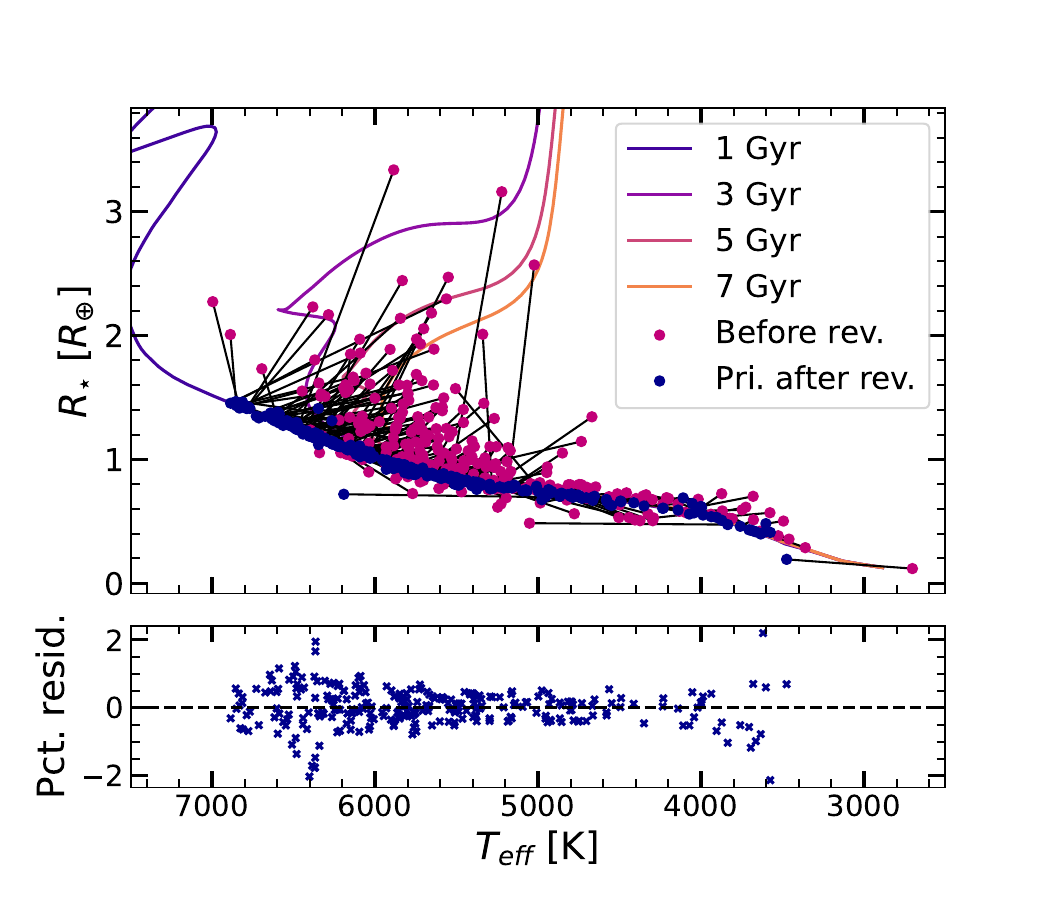}
    \caption{Radius versus \teff\ for all observed systems, showing both initial Kepler values from DR25 \citep[magenta points;][]{Berger2018} and the revised primary star values from this work (navy points), with the initial and revised values for each system connected by a black line. The underlying curves are MIST isochrones at various ages. The lower panel indicates the percent residual between the 1 Gyr evolutionary model and the final primary star radius for accepted systems.}
    \label{fig:hrd}
\end{figure}

\begin{deluxetable*}{CCCCCCCCCC}
\tablecaption{Stellar parameter fit results for all KOIs in our sample \label{tab:star_table}}
\setcounter{table}{3}
\tablecolumns{10}
\tablewidth{0pt}
\tablehead{
\colhead{KOI} & \colhead{$T_{{\rm eff},1}$} & \colhead{$T_{{\rm eff},2}$} & \colhead{$T_{Kepler}$} & \colhead{$R_{1}$} & \colhead{$R_{2}/R_{1}$} & \colhead{$R_{Kepler}$} & \colhead{f$_{corr, p}$} & \colhead{f$_{corr, s}$} & \colhead{Source}\\
\colhead{} & \colhead{(K)} & \colhead{(K)} & \colhead{(K)} & \colhead{\rsun} & \colhead{} & \colhead{\rsun} & \colhead{} & \colhead{} & \colhead{}
}
\startdata
42 & $6640^{+17}_{-14}$ & $4771^{+43}_{-38}$ & $6306\pm126$ & $1.376^{+0.009}_{-0.009}$ & $0.516^{+0.005}_{-0.005}$ & $1.509\pm0.063$ & $0.94^{+0.04}_{-0.04}$ & $2.01^{+0.10}_{-0.09}$ & 1\\
112 & $6256^{+65}_{-61}$ & $5506^{+87}_{-86}$ & $5833\pm105$ & $1.124^{+0.039}_{-0.037}$ & $0.753^{+0.009}_{-0.009}$ & $1.022\pm0.143$ & $1.27^{+0.21}_{-0.17}$ & $1.67^{+0.29}_{-0.21}$ & 2\\
162 & $6148^{+104}_{-118}$ & $5622^{+164}_{-188}$ & $5788\pm116$ & $1.078^{+0.054}_{-0.053}$ & $0.811^{+0.060}_{-0.054}$ & $1.124\pm0.164$ & $1.17^{+0.20}_{-0.17}$ & $1.42^{+0.26}_{-0.20}$ & 2\\
163 & $5290^{+54}_{-55}$ & $5015^{+52}_{-53}$ & $5078\pm101$ & $0.763^{+0.008}_{-0.008}$ & $0.956^{+0.004}_{-0.004}$ & $0.766\pm0.052$ & $1.30^{+0.09}_{-0.09}$ & $1.48^{+0.11}_{-0.09}$ & 2\\
165 & $5508^{+87}_{-94}$ & $4432^{+90}_{-79}$ & $5211\pm103$ & $0.850^{+0.031}_{-0.028}$ & $0.772^{+0.007}_{-0.007}$ & $0.799\pm0.060$ & $1.16^{+0.11}_{-0.09}$ & $2.07^{+0.18}_{-0.17}$ & 2\\
\enddata
\tablecomments{This table presents the revised stellar temperatures and radii from this work, as well as the \kepler\ values from \citet{Berger2018} when available (flagged with source = 1), or otherwise from \citet{Mathur2017} (flagged with source = 2); the composite stellar system properties; and the planetary radius correction factor if the primary or secondary star is the host. A truncated version of the table is shown here for reference, and the remainder of the table is available in machine-readable format from the journal.}
\end{deluxetable*}

\begin{figure*}
    \plottwo{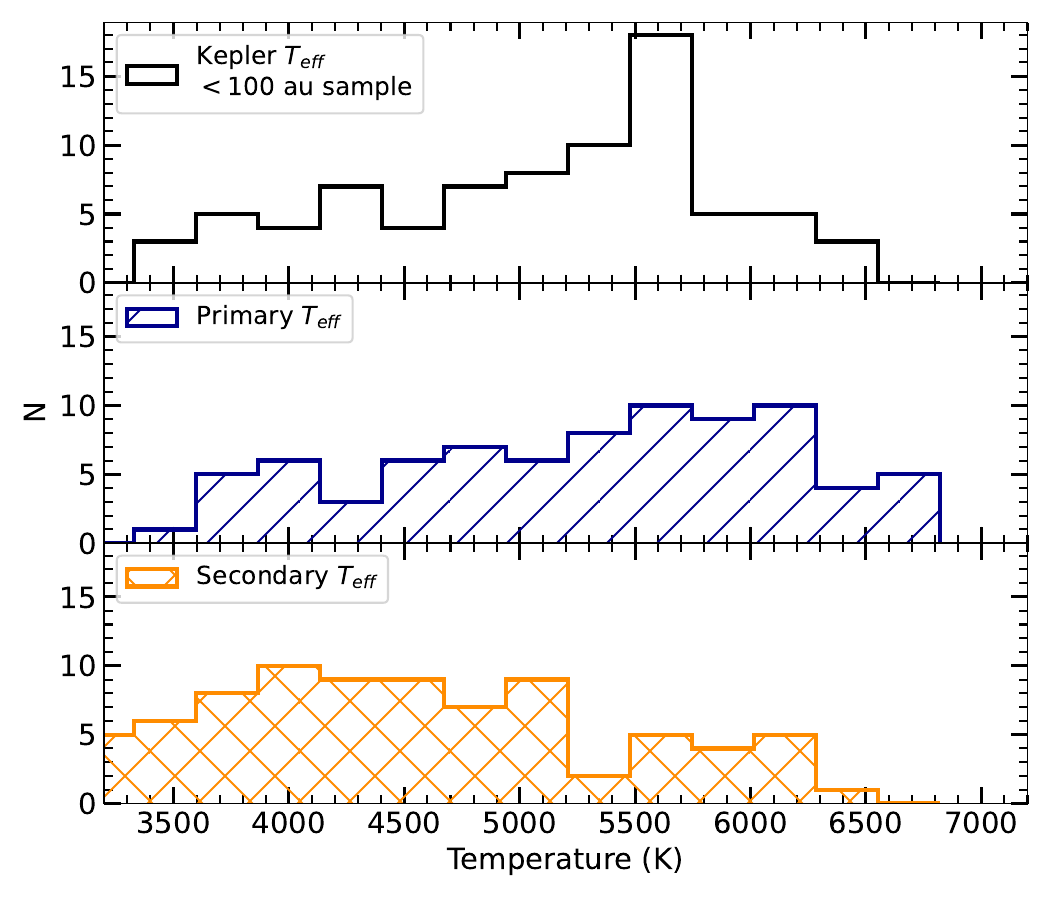}{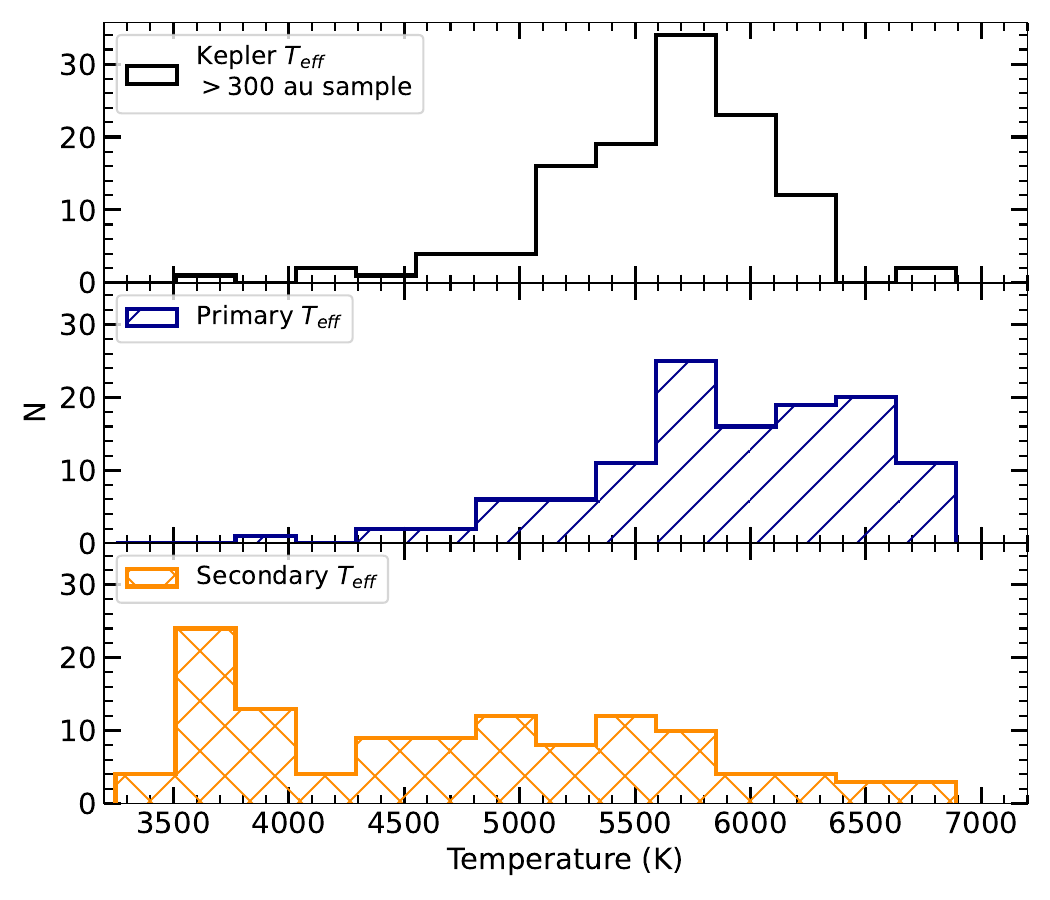}\\
    \includegraphics[width = 0.45\linewidth]{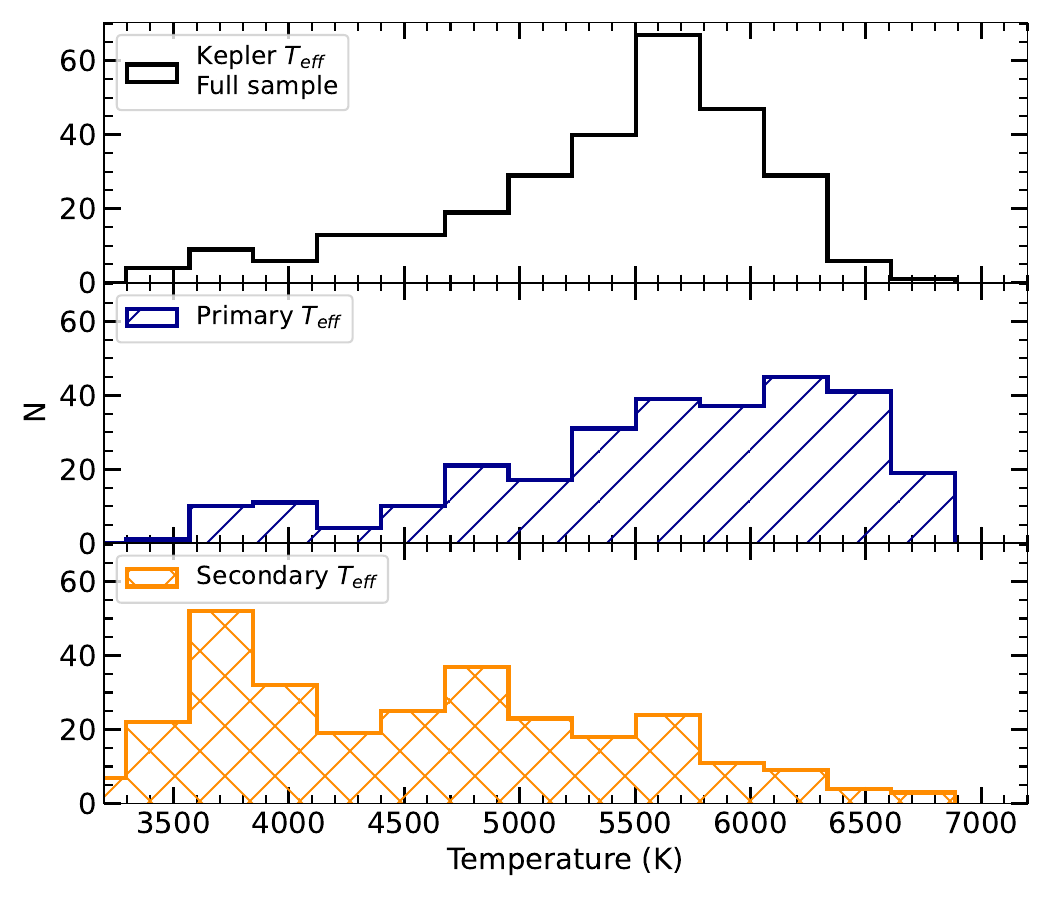}
    \caption{Histograms of the input (\citealt{Berger2018} or \citealt{Mathur2017}) stellar temperatures (black, top panel), revised primary star temperatures (blue, middle panel), and secondary star temperatures (orange, bottom panel). The three figures show the $\leq$100 au and $>$300 au samples on the top row, and the full sample below. The $\leq$100 au sample has more cool primary stars than the $>$300 au sample, likely because the small-separation binaries are systematically more nearby than the wider binaries, meaning that lower-mass stars are detectable.}
    \label{fig:teff_hist}
\end{figure*}

We imposed a Gaussian prior on the system parallax using the Gaia DR3 parallax and its associated error in most cases, and the ExoFOP Kepler DR25 distance if there was no published Gaia parallax. The majority of our sources (90\%) were included in Gaia. We placed a Gaussian prior with a spread of 5\% on the stellar radii using radii derived using the MESA Isochrones and Stellar Tracks (MIST) evolutionary models \citep{Paxton2011, Paxton2013, Paxton2015, Choi2016, Dotter2016} at an age of 1 Gyr. We imposed a Gaussian prior on the interstellar extinction using the Bayestar2019 3-dimensional dust map \citep{Green2019} implemented in the \texttt{dustmaps} package \footnote{\url{https://dustmaps.readthedocs.io/en/latest/index.html}}. Figure \ref{fig:hrd} shows the input versus output stellar radii and \teff\ from the \citet{Berger2018} (when available) or \citet{Mathur2017} catalogs versus this work. Approximately 60\% of our sample was analyzed in \citet{Berger2018}. 

The revised stellar parameters are summarized in Table \ref{tab:star_table} for all systems. The input and output stellar temperature distributions for the full sample and two subsamples ($\rho \leq 100$ au and $\rho > 300$ au) are shown in Figure \ref{fig:teff_hist}. The 100 au sample has more cool stars because close binaries are more difficult to detect at larger distances. The maximum distance for the $\leq$100 au sample is $\sim$1100 pc, while the maximum distance for the $>$300 au sample is $\sim$2500 pc. We tested our analysis with both samples truncated at a \teff\ lower limit of 5000 K and found that our results did not change significantly.

\subsection{Planet Parameter Revision}

\begin{deluxetable*}{CCCCCCCCCC}
\tablecaption{Planet parameter fit results for all planets in the binary KOI sample \label{tab:planet_table}}
\setcounter{table}{4}
\tablecolumns{10}
\tablewidth{0pt}
\tablehead{
\colhead{KOI} & \colhead{R$_{p, pri}$} & \colhead{R$_{p, sec}$} & \colhead{R$_{Kep}$} & \colhead{T$_{eq, pri}$} & \colhead{T$_{eq, sec}$} & \colhead{T$_{eq, Kep}$} & \colhead{S$_{pri}$} & \colhead{S$_{sec}$} & \colhead{S$_{Kep}$}\\
\colhead{} & \colhead{($R_{\earth}$)} & \colhead{($R_{\earth}$)} & \colhead{($R_{\earth}$)} & \colhead{(K)} & \colhead{(K)} & \colhead{(K)} & \colhead{(S$_{\earth}$)} & \colhead{(S$_{\earth}$)} & \colhead{(S$_{\earth}$)}
}
\startdata
42.01 & $2.28^{+0.14}_{-0.14}$ & $4.89^{+0.32}_{-0.33}$ & $2.43\pm0.12$ & $873^{+27}_{-27}$ & $451^{+15}_{-15}$ & $866$ & $148.51^{+-0.75}_{-6.32}$ & $15.80^{+0.50}_{-0.95}$ & $132.94\pm20.74$\\
112.01 & $3.55^{+0.72}_{-0.72}$ & $4.74^{+0.98}_{-0.98}$ & $2.75\pm0.39$ & $571^{+44}_{-44}$ & $436^{+35}_{-35}$ & $503$ & $22.10^{+1.85}_{-1.95}$ & $8.59^{+0.83}_{-0.82}$ & $15.17\pm6.00$\\
112.02 & $1.49^{+0.29}_{-0.30}$ & $1.98^{+0.38}_{-0.40}$ & $1.16\pm0.17$ & $1371^{+101}_{-104}$ & $1047^{+78}_{-80}$ & $1206$ & $729.36^{+61.17}_{-64.29}$ & $283.49^{+27.43}_{-27.20}$ & $500.46\pm197.96$\\
162.01 & $3.72^{+0.86}_{-0.83}$ & $4.50^{+0.93}_{-0.92}$ & $3.11\pm0.45$ & $848^{+67}_{-68}$ & $698^{+62}_{-66}$ & $807$ & $106.94^{+15.00}_{-15.30}$ & $55.45^{+11.28}_{-11.62}$ & $100.09\pm42.19$\\
163.01 & $2.71^{+0.26}_{-0.26}$ & $3.09^{+0.30}_{-0.29}$ & $2.08\pm0.14$ & $676^{+28}_{-27}$ & $626^{+25}_{-25}$ & $649$ & $51.10^{+2.71}_{-2.79}$ & $38.20^{+2.17}_{-2.08}$ & $42.07\pm9.84$\\
\enddata
\tablecomments{The revised and \citet{Thompson2018} planetary radii, instellations, and equilibrium temperatures. The entire sample of planets is included, although we only used systems with $0.5 \leq R_{p} \leq 4$ \rearth\ for the remainder of the analysis. Five rows of the table are shown here for reference, and the full table is available in the supplementary material.}
\end{deluxetable*}

We used the revised stellar parameters to recalculate the planet parameters. In general, the revised primary star radii were larger than the previous assumed stellar radius and the effective temperatures were higher. Additionally, the flux dilution from the secondary star \citep{Ciardi2015} meant that the measured planetary radii were systematically smaller than the true planet radii. In the case of a single star,

\begin{equation}
\delta \propto \left( \frac{R_{p}}{R_{\star}} \right)^{2}
\end{equation}
where $\delta$ is the transit depth, and $R_{p}$ and $R_{\star}$ are the planetary and stellar radii, respectively. For a binary, the effective (or inferred) stellar radius is not the true radius of the planet host, and there is an additional flux dilution term because of the companion star, meaning that the transit depth becomes

\begin{equation}
\delta \propto \frac{F_{\rm host}}{F_{\rm total}} \left( \frac{R_{p}}{R_{\star}} \right)^{2}
\end{equation}
where $F_{\rm host}$ and $F_{\rm total}$ are the host star flux and the total system flux in the Kepler bandpass, respectively. This leads to the assertion that 

\begin{equation}
    R_{p, corr} = R_{p} \sqrt{\frac{F_{\rm tot}}{F_{\rm host}}}
\end{equation}

Where the flux ratio between the total system flux and the host flux is the multiplicative ``Planet Radius Correction Factor'' \citep[PRCF;][]{Furlan2017} with which to correct the apparent (observed) planet radius. The correction factors for the primary and secondary host star cases have a similar form in the case of a two-star system with no additional flux components:

\begin{equation}
    f_{\rm pri} = \frac{R_{\star, {\rm pri}}}{R_{\star, {\rm total}}} \sqrt{1 + 10^{-0.4 \Delta m_{Kep}}}
\end{equation}
\begin{equation}
    f_{\rm sec} = \frac{R_{\star, {\rm sec}}}{R_{\star, {\rm total}}} \sqrt{1 + 10^{0.4\Delta m_{Kep}}} \\
\end{equation}

\begin{figure}
    \plotone{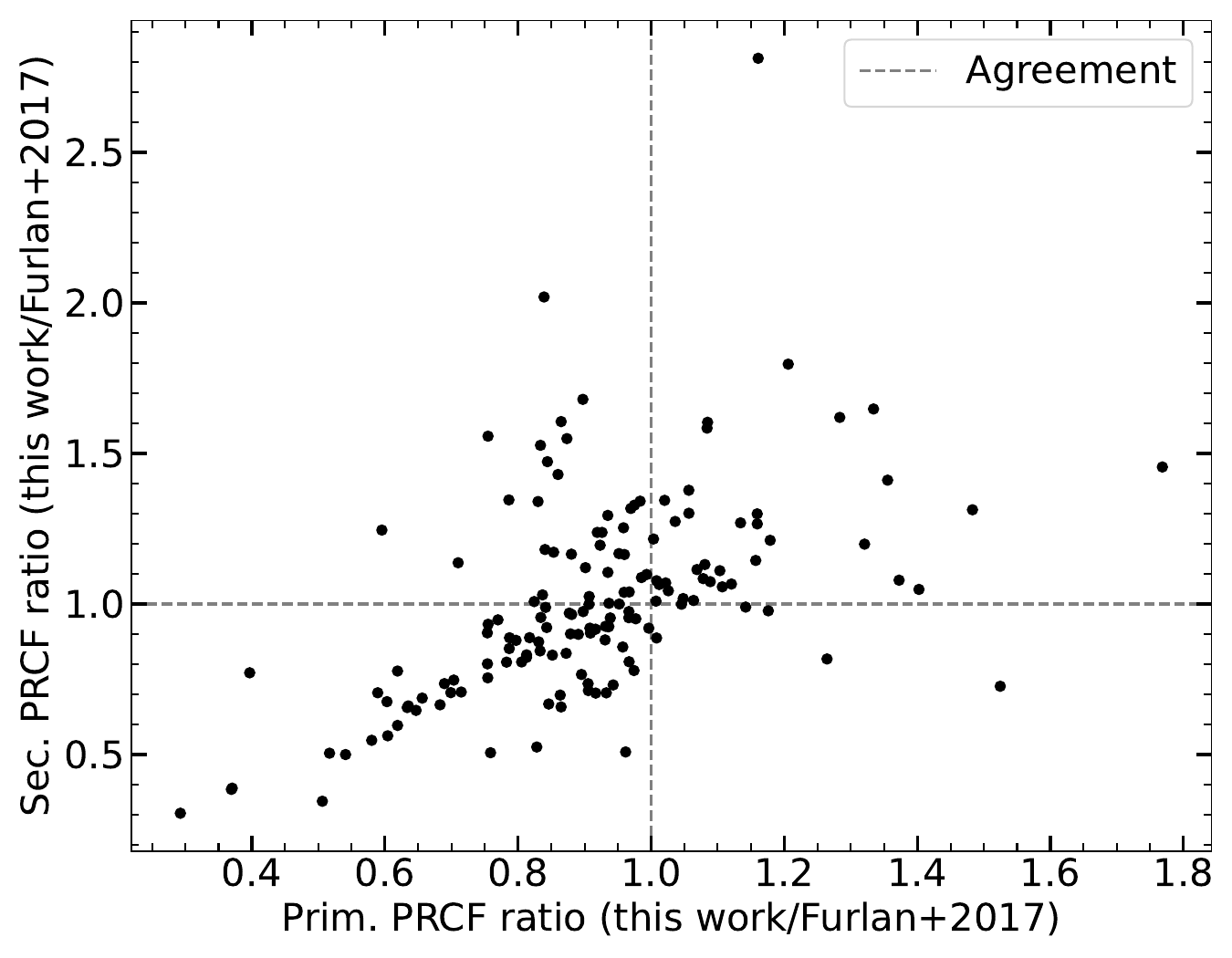}
    \caption{The fractional difference between the primary and secondary PRCF calculated in this work versus the PRCF of \citet{Furlan2017}. The vertical and horizontal lines at 1 denote perfect agreement. The majority of systems agree fairly closely between this work and \citet{Furlan2017}, but there are many outliers with disagreement up to 60\% in primary PRCF, and even more in secondary PRCF. This is likely because of better distances from Gaia and stellar parameter revision improving stellar radius estimates.}
    \label{fig:prcf_compare}
\end{figure}

as derived in \citet{Furlan2017}. In contrast to \citet{Furlan2017}, we did not assume that the primary star radius was equal to the inferred Kepler radius from Kepler DR25 \citep{Thompson2018}, but instead corrected for the revised primary radius based on our stellar parameter revision. For systems included in both \citet{Furlan2017} and this work, we compared the calculated PRCF from that work and found that they typically agreed within 20\%, but there was substantial scatter with some outliers disagreeing by up to 60\% after incorporating the primary star radii (Figure \ref{fig:prcf_compare}). This is likely the result of new distances from Gaia improving our estimates of the stellar radii and the binary star analysis process resulting in different measured stellar radii. 

The semi-major axis and incident flux of the planets also changed because of the altered stellar radius (and thus the inferred stellar mass). We calculated the new semimajor axis $a$ using Kepler's third law, and calculated the new insolation flux $S$ based on the new semimajor axis and stellar luminosity. In both cases we used the new inferred stellar mass and luminosity using the MIST evolutionary models at an age of 1 Gyr given our measured stellar \teff. The revised planetary radii, insolation fluxes, and equilibrium temperatures are presented in Table \ref{tab:planet_table}, along with the Kepler DR25 values from \citet{Thompson2018}.

\section{The Radius Distribution of Planets in Binary Systems}
We assembled a sample of binary systems hosting planets with apparent radii $0.5\ R_{\oplus} \leq R_{p} \leq 4\ R_{\oplus}$ which we observed using HET/LRS2 and Keck/NIRC2, then analyzed using the new data as well as data from the literature. Then, we revised the planetary properties to correct for the presence of the secondary star. To assemble a final sample, we selected planets with $P \leq 100$\,d and fractional radius error $\sigma_{R_{p}} < 25\%$. We chose the period cutoff to restrict the sample to the regime where Kepler was most complete, and where other efforts have truncated their planet period distributions when studying the radius gap \citep[e.g.,][]{Fulton2017}. As discussed in Section \ref{sec:sample}, we removed planets with a large radius error to ensure that the original transit fit was high-quality. We were left with a total of 207 systems hosting 283 planets. Along with the full sample (marginalizing over binary separation) we also explored two subsamples: systems with binary projected separation $\rho \leq 100$ au ($N_{p} = 100,\ N_{\star} = 65$), and systems with binary projected separations $\rho > 300$ au ($N_{p} = 94,\ N_{\star} = 73$). These cuts were chosen to include the regime where multiplicity is expected to have the largest impact (the $\leq$100 au sample) or to have a minimal impact (the $>$300 au sample).

\subsection{The Planetary Radius Distribution and Approximating Detection Sensitivity}

\begin{figure}
    \plotone{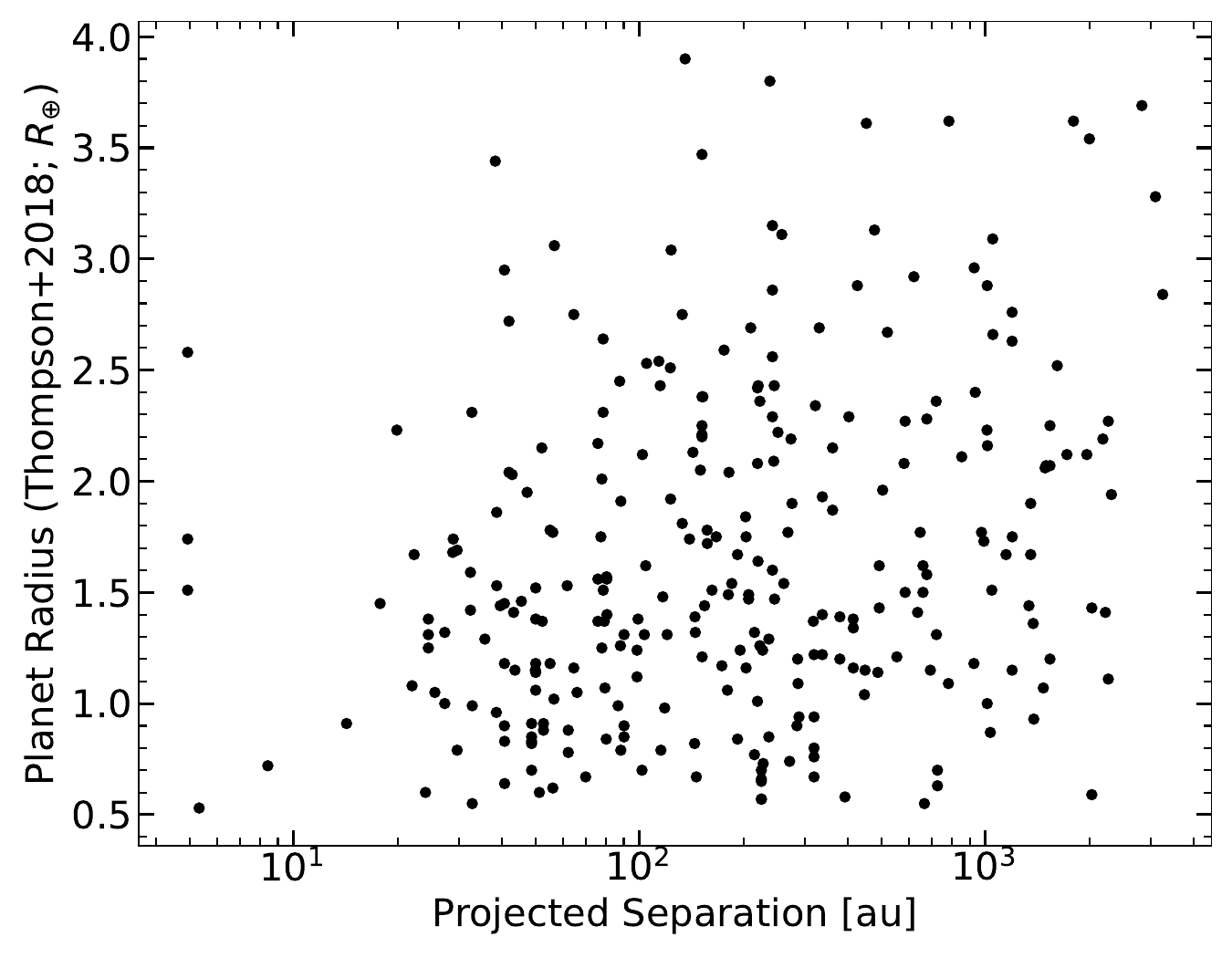}
    \caption{Planet radius from \citet{Thompson2018} plotted against projected physical separation. Before any parameter revision, it is already apparent that there may be a dearth of larger planets in small-separation binaries.}
        \label{fig:a_vs_radius}
\end{figure}

Before assessing the revised planetary radii, we wished to examine the input planetary radius distribution as a function of projected separation. Figure \ref{fig:a_vs_radius} shows the planetary radii from \citet{Thompson2018} plotted against the projected separation. Even before any planetary radius revision, it is apparent that the number of large planets decreases substantially as the binary separation decreases, with an upper envelope in planet size becoming apparent below $\sim 100$ au binary separation. This effect was also noted by \citet{Hirsch2021}.

\begin{figure*}
    \plottwo{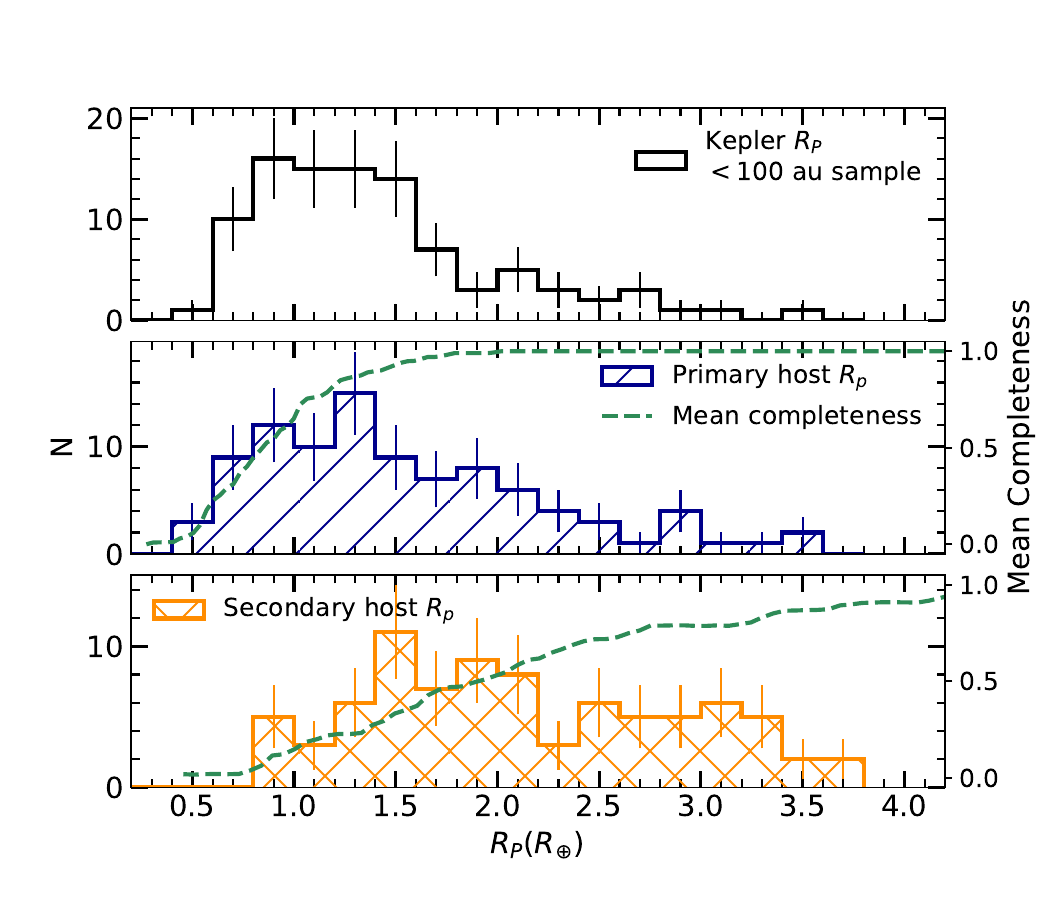}{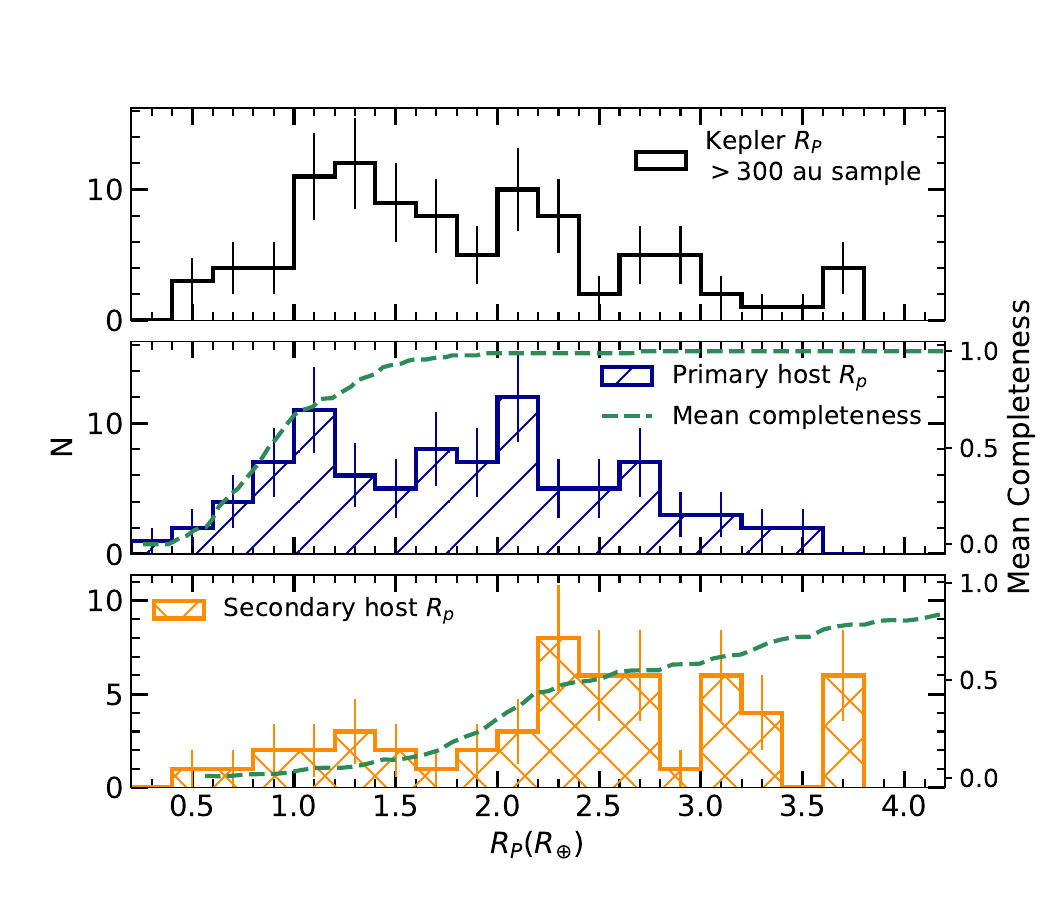}\\
    \includegraphics[width = 0.45\linewidth]{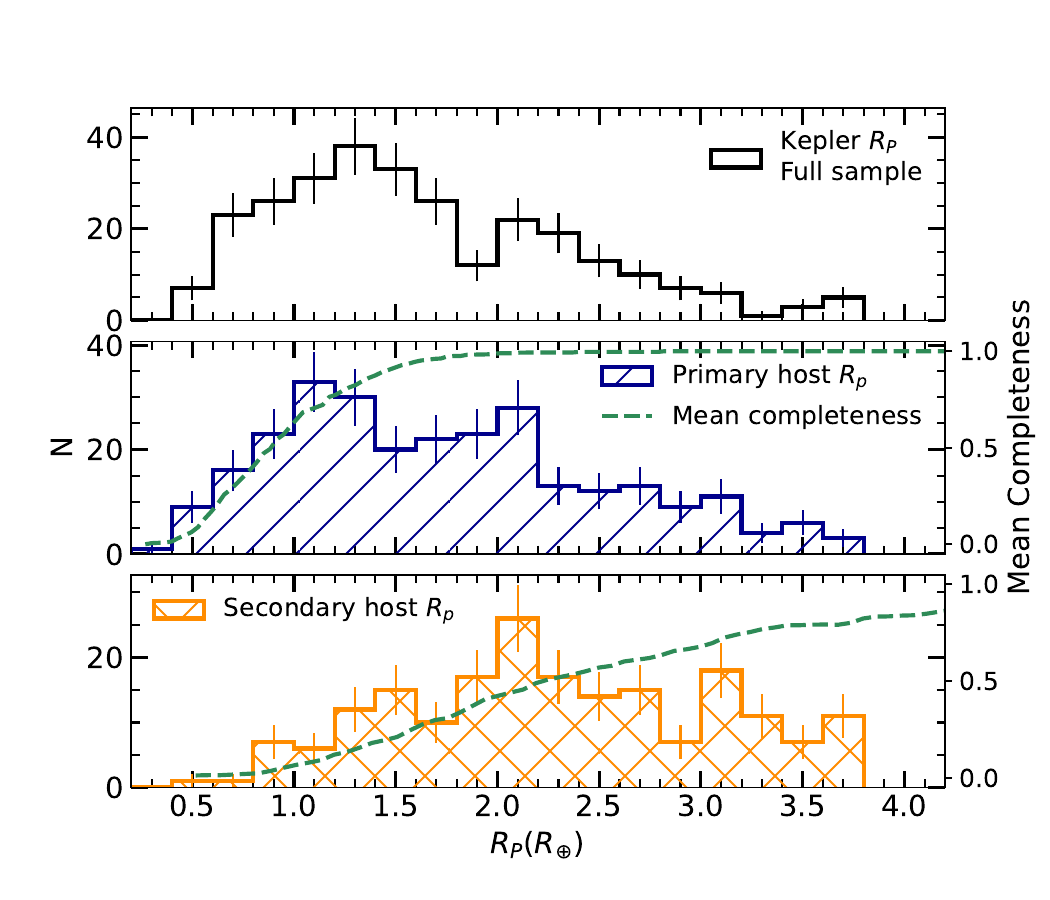}
    \caption{Histograms of the planet radius distribution using the input \citet{Thompson2018} Kepler DR25 values (top; black), assuming all systems are around the primary star (middle; blue) and all systems are around the secondary star (bottom; orange). The top row shows the $\rho \leq 100$ au sample (left) and the $\rho > 300$ au sample (right), and the full sample is shown on the bottom row. The dashed green curve shows the median completeness of all systems, described in the text. The full sample does not show evidence for a statistically significant gap, agreeing with \citet{Sullivan2023}, while the $\leq$100 au sample has few sub-Neptunes. The $>$300 au sample shows a radius gap that appears to be shifted to smaller planetary radii. The revised planetary radius histograms have longer tails toward larger radii compared to the input samples because of the planetary radius bias caused by binaries.}
    \label{fig:radius_hist}
\end{figure*}

Figure \ref{fig:radius_hist} shows the input $R_{p}$ distribution from \citet{Thompson2018}, as well as the revised radius distributions if all planets are around the primary or secondary star. We chose to use the values from \citet{Thompson2018} to ensure homogeneity of the sample, instead of mixing catalogs. The revised radius distributions also have mean completeness curves plotted, which we calculated in the following way: First, for each individual system we calculated a completeness curve by generating 1500 random planetary radii, then calculating a transit depth as $\delta = (R_{p}/R_{\star})^{2}$. We imposed a photometric noise value using the transit duration and the $\sigma_{CDPP}$ values for each system, available from the Exoplanet Archive. We took the measured planet period from the Exoplanet Archive, then calculated a new semimajor axis assuming a primary or secondary host and Kepler's third law:

\begin{equation}
    \frac{a}{\rm au} = \left(\frac{P}{\rm yr}\right)^{2} \left(\frac{M_{\star}}{M_{\odot}}\right)^{1/3}.
\end{equation}

Following \citet{Winn2010}, we calculated the transit duration as 

\begin{equation}
    t_{dur} = \frac{P}{\pi} \arcsin{\left(\frac{R_{\star} + R_{p}}{a}\right)}.
\end{equation}

Finally, from \citet{Batalha2013}, the S/N of the transit is

\begin{equation}
   {\rm S/N} = \left(\frac{\delta}{\sigma_{\rm SDCC}}\right)\sqrt{N_{transit}},
\end{equation}

where we calculated the number of observed transits from the recorded time baseline (taken from the Exoplanet Archive) and the planet period. Then, we imposed a detection cutoff of S/N = 7. We calculated a sensitivity curve for each system assuming either a primary or secondary star host using the planet radius correction factor, then took the mean at each planetary radius to calculate the mean curve for all systems. These curves are intended to be approximations for illustrative purposes. A full injection-recovery analysis for the detection sensitivity is outside the scope of this work. 

Figure \ref{fig:radius_hist} shows that we are nearly 100\% sensitive to planets hosted by primary stars above $\sim 1.4$ \rearth, while the sensitivity to planets in secondary star hosts is much lower because of the increased flux dilution from the primary star. This result supports our decision to assume that all planets are around the primary stars for our subsequent analysis, since only a small fraction of the planets around secondary stars would be detected. We tested our assumption later in the analysis by also checking mixtures of 90\%/10\% and 50\%/50\% primary/secondary host samples, and found that our results did not change significantly. Determining the host star for planets in binaries is a challenging task in itself, so we defer future discussion of planet host stars and detection efficiency for primary versus secondary stars to future work.

In Figure \ref{fig:radius_hist} we show the full sample (all binary separations) as well as two subsamples, one with binary projected separations $\rho \leq 100$ au and another with $\rho > 300$ au. In each of the subsamples, the input sample from \citet{Thompson2018} shows demographic differences that are exacerbated after the planetary radius revision. For the $\rho \leq 100$ au sample (top left panel of Figure \ref{fig:radius_hist}), there is no apparent radius gap. Instead, there is a single peak around the location of the super-Earth peak in single stars, about 1.3 \rearth.

For the $\rho > 300$ au sample (top right panel of Figure \ref{fig:radius_hist}), the input population may show a radius gap, as does the output distribution. In the output distribution the gap appears to be shifted to smaller radii than in the input distribution. Again, the apparent difference between the input and output (revised) populations is not large, but the distribution gains a longer tail toward the sub-Neptunes. 

The full sample of all stars with projected separations ranging from 4-2000 au is shown in the bottom panel of Figure \ref{fig:radius_hist}. Consistent with \citet{Sullivan2023}, the full sample does not show evidence for a statistically significant radius gap, even though a gap may be visible in the input population of planets in binaries without revised radii. We define statistical significance as a reduction in the number of planets that is larger than 1$\sigma$ of the Poisson error in the different planetary radius bins. The sub-samples support the suggestion from \citet{Sullivan2023} that the radius distribution changes as a function of binary separation, and may obscure any presence of a gap in the full population. This implies that a population of undetected binaries in a nominally single star sample may indeed contaminate the radius gap, making it appear shallower than it truly is for single stars. 

\subsection{Comparing the Radius Distribution of Planets in Binaries vs. Single Stars}

\begin{figure*}
    \plottwo{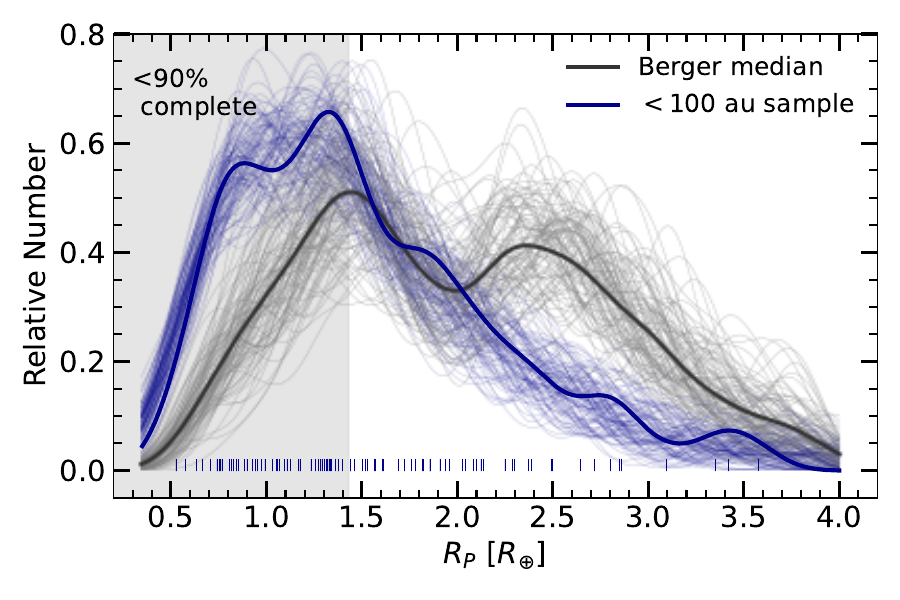}{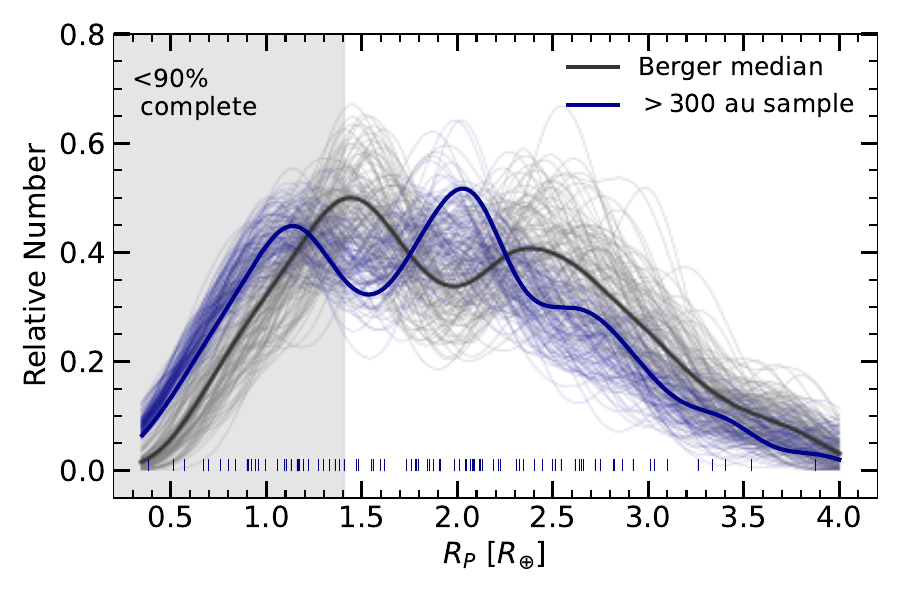}\\
    \includegraphics[width = 0.5\linewidth]{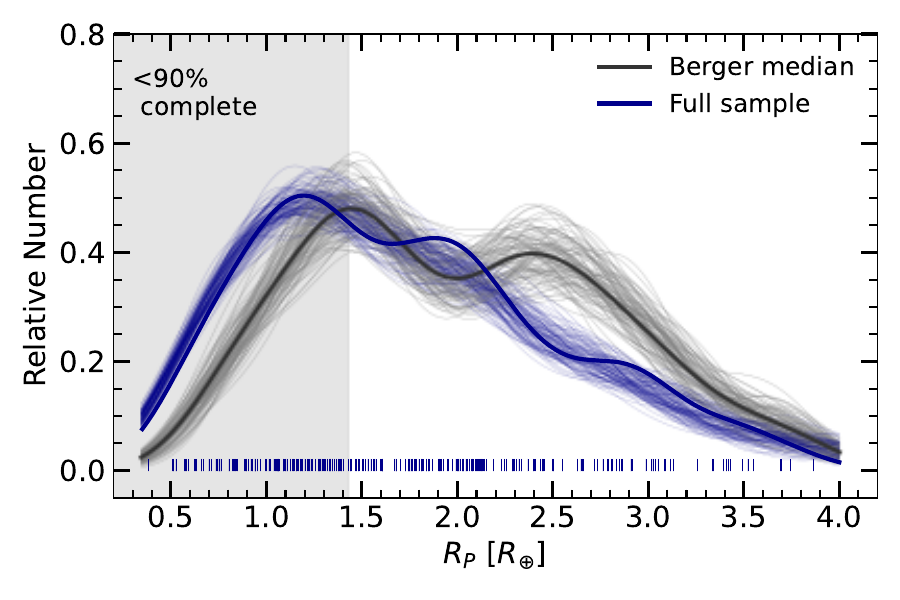}
    \caption{One dimensional normalized kernel density estimates (KDEs) of the planet radius distribution for the $\rho \leq 100$ (top left), $\rho > 300$ (top right), and full samples (bottom row), shown in purple, and the \citet{Berger2020} single star sample, shown in black. The radius regime where a given planetary radius could be detected by fewer than 90\% of the systems is shaded gray. The thin lines behind each KDE are random draws from each sample, with the binary planets randomly perturbed based on their error and the \citet{Berger2020} sample redrawn with the same sample size as the binaries. The small-separation sample shows a smaller relative number of sub-Neptunes, while the wider-separation sample resembles the \citet{Berger2020} single-star distribution with a possible shift in the location of the radius gap.}
    \label{fig:radius_KDE}
\end{figure*}

From our sensitivity calculations we found that we are sensitive to most small planets around primary stars, but very few small planets around the secondary stars. This supports our decision to focus on primary star hosts for the remainder of the analysis. To explore the radius distribution of planets in binaries more deeply and compare directly to a single-star sample, we calculated a 1-D Kernel Density Estimate (KDE) for each subsample, as well as for a sample from \citet{Berger2020} after selecting comparison systems using the same criteria we imposed on our input sample ($P \leq 100$ d, $\sigma_{R_{p}} < 25$\%). A KDE is similar to a histogram, but rather than separating points into discrete boxes, it smooths the distribution using a specified kernel for each measured point. The advantage of a KDE rather than a histogram is that the structure of the distribution is less dependent on binning. We chose the kernel size of the KDE to be the median radius uncertainty in each sample, $\sim$0.12 \rearth\ in all cases. To perform any statistical tests, we used the measured planet radii directly, rather than the KDE curves.

The resulting KDEs are shown in Figure \ref{fig:radius_KDE} with the $<$90\% mean sensitivity region shaded in gray (i.e., the radii where a planet of that size would be detected in fewer than 90\% of the systems in the sample). We made 100 draws from each sample using a random selection of eligible stars from \citet{Berger2020} and random draws from the error distribution for the binaries to demonstrate the scatter in each distribution. The thick solid lines are calculated using the full eligible single star sample from \citet{Berger2020} and the preferred radius values from the binary sample. 

\begin{deluxetable}{CCCC}
\tablecaption{Sample comparison statistics \label{tab:stats}}
\setcounter{table}{5}
\tablecolumns{4}
\tablewidth{0pt}
\tablehead{
\colhead{Sample} & \colhead{K-S test} & \colhead{A-D test} & \colhead{Ratio}
}
\startdata
$\rho \leq 100$\ {\rm au} & $<$0.001 & $<$0.001 & 2.47$\pm$0.31\\
$\rho > 300$\ {\rm au} &  $> 0.25$ & 0.08 & 0.92$\pm$0.28\\
{\rm All} & 0.006 & 0.004 & 1.38$\pm$0.16\\
\enddata
\tablecomments{Table values are p-values. We consider a value of $p < 0.05$ to indicate a statistically significant difference between samples. The ratio is the relative number of super-Earths ($1 \leq R_{p} \leq 1.7$ \rearth) versus sub-Neptunes ($2.2 \leq R_{p} \leq 3.5$ \rearth). The errorbars were calculated using Poisson statistics.}
\end{deluxetable}

Assuming all planets were hosted by the primary star, we calculated the similarity between each subsample of planetary radii and the \citet{Berger2020} sample using two-sample Kolmogorov-Smirnov (K-S) and Anderson-Darling (A-D) tests as implemented in scipy \citep{Virtanen2020}. K-S and A-D tests are sensitive to the mean and tails of the distribution, respectively. The results from both tests are presented in Table \ref{tab:stats} for each sample. We found that both the $\rho \leq 100$ au sample and the full sample are significantly different from the single stars using both the K-S and A-D tests, while the $\rho > 300$ au sample is consistent with the single stars, with p-values from both tests $p > 0.05$, which indicate no statistically significant difference between the two samples. However, the location of the radius gap appears to be slightly shifted to smaller planetary radii in both the 300 au sample and the full sample relative to the single stars. 

We also calculated the ratio between the number of super-Earths ($1 \leq R_{p} \leq 1.7$ \rearth) and sub-Neptunes ($2.2 \leq R_{p} \leq 3.5$ \rearth) for each sample (Table \ref{tab:stats}). Because we did not calculate occurrence rates for our sample these values are a lower limit, because the number of super-Earths is likely underestimated because of limited completeness at small radii. This lower limit shows that there are at least 2.47$\pm$0.31 super-Earths for each sub-Neptune in the $\leq$100 au sample, while the ratio is 0.92$\pm$0.28 and 1.38$\pm$0.16 for the $>$300 au sample and the full sample, respectively. Although these are lower limits, sensitivity between the samples is approximately equivalent, suggesting that there are at least twice as many super-Earths relative to sub-Neptunes in the $\leq$100 au sample as in the other samples. These limits support our result that the sub-Neptunes are suppressed relative to the super-Earths in the $\leq$100 au sample. However, occurrence rate calculations and determining which star in each binary hosts the planet are both crucial for validating this result.

\begin{figure}
    \plotone{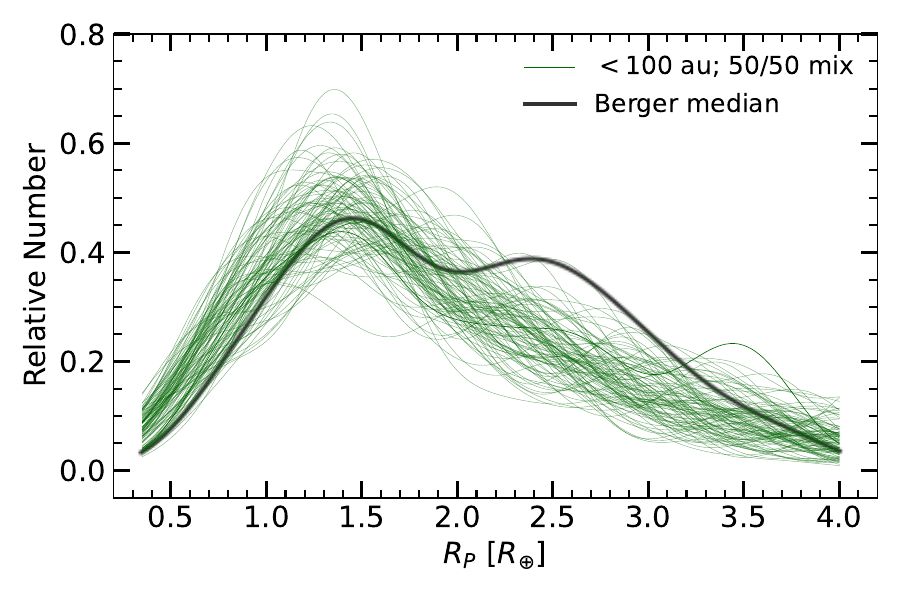}
    \caption{KDEs of 100 random draws from the planet radius posteriors from the $\leq100$ au sample, with each draw composed of a 50/50 mixture of primary and secondary star hosts (in green), compared to the \citet{Berger2020} radius distribution in black. Although the addition of secondary star hosts increases the spread and occasionally leads to a planet radius peak at larger radii, the overall structure of the KDE is not impacted by the inclusion of planets around the secondaries.}
    \label{fig:5050}
\end{figure}

Similar to the histograms in Figure \ref{fig:radius_hist}, it is apparent in Figure \ref{fig:radius_KDE} that there are differences in the planet radius distribution between subsamples, as well as differences between the binary star and single star distributions. To test the robustness of our results, we performed tests exploring the impacts of slightly altered analyses on our observed radius distributions. First, when calculating the KDEs, we tested imposing a 10\% mixture of secondary star hosts and a 90\% mixture of primary star hosts, to account for potential contamination by secondary star hosts and explore whether the addition of secondary star hosts would impact the observed demographics. We also tested a 50/50 mixture, an example of which is shown in Figure \ref{fig:5050} for the $\leq$100 au sample. We found that including secondary star hosts increased scatter but did not significantly change the observed structure. We performed a K-S test between the \citet{Berger2020} median radius distribution and each test of the 50/50 mixture and found that the median p-value from the K-S test was $p < 0.04$, indicating that the single star versus binary star distributions remained significantly different using our stated metric for statistical significance of $p < 0.05$.

\begin{figure}
    \plotone{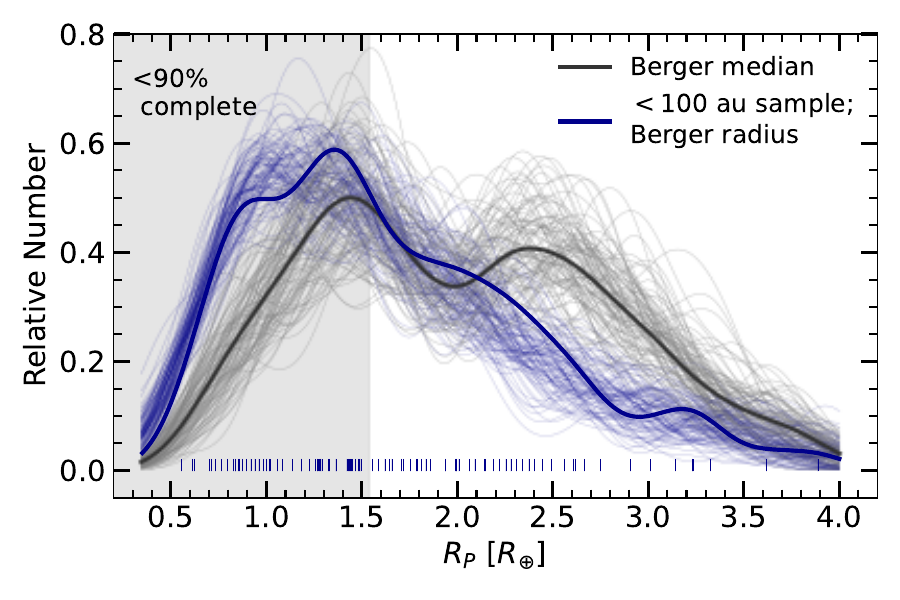}
    \caption{The same as Figure \ref{fig:radius_KDE} for the $\leq$100 au sample, except with the primary star radius assumed to be the stellar radius from \citet{Berger2018} or \citet{Mathur2017}, instead of the result from our reanalysis using a 1 Gyr isochrone prior on the ages. The result does not change significantly between the two different stellar radius assumptions, as discussed in the text.}
    \label{fig:berger_radii}
\end{figure}

To check that our choice of a 1 Gyr isochrone to impose the radius priors was not biasing our result, we also ran our analysis using the \citet{Berger2018} and \citet{Mathur2017} radii as the primary star radius, and did not find significant differences, as shown in Figure \ref{fig:berger_radii}. The A-D and K-S tests found p-values of $p_{K-S} <  0.001$ and $p_{A-D} < 0.001$, and the ratio between the super-Earths and the sub-Neptunes was $SE/SN = 1.75 \pm 0.28$, which is 1-$\sigma$ discrepant with the result from the analysis performed with our revised stellar radii.

\section{Discussion}
We have revised the radii of a sample of 283 small transiting planets with $R_{p} \leq 4$ \rearth\ and $P \leq 100$ d in 207 binary star systems with physical separations ranging from $\sim 4-2000$ au. We explored the radius distribution of the planets as a function of binary star separation, including examining the radius distribution in the $R_{p}-S_{\oplus}$ plane. We found that the planetary radius distributions in close ($\rho \leq 100$ au) and wide ($\rho > 300$ au) binaries are distinct from each other, and that the close binary planetary radius distribution is also significantly different than the planet radius distribution for single stars. In this section we contextualize our results within the literature. 

In the $\leq$100 au sample, there appears to be a significantly smaller relative number of sub-Neptunes when compared to the single-star radius distribution, as seen in Figure \ref{fig:radius_KDE}, which compares the $\leq$100 au sample to the single-star sample of \citet{Berger2020}. The lack of sub-Neptunes is also apparent in both the input and output radius distributions in Figure \ref{fig:radius_hist}. 

One possible explanation for the lack of sub-Neptunes is a detection bias. However, the detection curves estimated in Figure \ref{fig:radius_hist} show that our sensitivity to planets is high above $\sim$ 1.5 \rearth. We should also be progressively less sensitive to smaller planets, meaning that an overabundance of super-Earths and underabundance of sub-Neptunes is the inverse of what we would expect if our results were significantly impacted by a detection bias. It is possible that there is a detection bias against detecting super-Earths in the $>$300 au sample, for example because the wider binaries can be resolved at larger distances, where smaller planets become more difficult to detect. However, this potential effect does not explain the structural differences between the $\leq$100 au and $>$300 au samples (i.e., the unimodal distribution and lack of a radius gap in the $\leq$100 au sample). For example, it is unlikely that the $\leq$100 au and $>$300 au samples have substantial differences between how many planets (or how many sub-Neptunes vs. super-Earths) are hosted by the secondary star, meaning that we expect the structural differences between the $\leq$100 au and $>$300 au samples to be robust against detection bias.

Another possible bias is that a significant number of the planets could in fact be sub-Neptunes around the secondary star that we mischaracterized by assuming they were around the primary star. However, there is a bias against detecting transiting exoplanets around secondary stars because of flux dilution, meaning that it is unlikely that a large fraction of the planets in our sample are orbiting the secondary star. This is supported by the sensitivity curves we calculated in Figure \ref{fig:radius_hist}, where the sensitivity to planets with secondary star hosts is much lower than those around primary star hosts. Furthermore, we tested mixtures of 90/10 and 50/50 primary versus secondary star planet hosts (rather than our fiducial 100\% primary host assumption) and our results did not change significantly. However, future work will investigate the primary host star assumption and provide further clarity on the impacts of possible secondary star planet hosts.

Instead of a detection bias, there may be a physical explanation for the apparent lack of sub-Neptunes in close binary star systems. Many protoplanetary disks in binary systems have reduced lifetimes, dissipating about 10 times faster than their counterparts around single stars \citep{Cieza2009, Kraus2012, Barenfeld2019}. Intuitively, this reduced disk lifetime suggests that a planetary core may not have enough time to accumulate and subsequently retain a gaseous atmosphere. 

Alternatively, not all disks in close binaries appear to dissipate extremely rapidly, with about one third of the disks surviving as long as those in single stars \citep[e.g.,][]{Kraus2012}, suggesting that the timescale for planet formation in close binaries may not always occur on a shorter timescale. However, even in these longer-lived binary disks, there could be disruptions to the disk via disk truncation and dynamical disruption and stirring from the outer companion \citep[e.g.,][]{Artymowicz1994}. For example, the truncated outer disk could reduce the mass budget of solids or change the balance between ice and rock in the disk, impacting final planet size and composition \citep[e.g.,][]{Drazkowska2023}. Studies of disk structure in close binaries \citep[e.g.,][]{Sullivan2022a, Tang2023} and additional searches for planets in young binary systems will illustrate the relative importance of disk lifetime, disk structure, dynamical disruption, and disk radius on planet occurrence and properties.

\citet{Rogers2023} found that the final atmospheric fraction is also dependent on planetary core mass, suggesting that the smaller cores that could be formed over the shorter disk lifetime \citep{Drazkowska2023} would be more susceptible to atmospheric mass loss. This would naturally lead to a shift in the peak location of the super-Earth radius distribution, which falls around 1.3 \rearth\ for single stars. However, we do not have sufficient sensitivity to small planets in binaries to explore this prediction; in fact, sensitivity to small planets in our sample substantially declines below $\sim$ 1.4 \rearth\ (as noted for other samples; \citealt{Lester2021}), meaning that we cannot confidently state where the peak of the super-Earth radius distribution falls. Future efforts that are more sensitive to small planets should investigate whether the super-Earth peak shifts as a function of binary separation. Calculations of the occurrence rate of planets in binaries rather than the observed radius distribution will help clarify the expected location of the peak of the super-Earth radius distribution and potentially establish whether smaller planetary cores may contribute to the lack of sub-Neptunes in close binary stars. 

Another intriguing feature of the $\leq$100 au sample is the presence of sub-Neptunes without an apparent radius gap. The radius gap is expected to be an intrinsic feature of exoplanet populations, produced during the first $\sim$ Gyr (or longer) via atmospheric mass loss \citep{Berger2020, David2021, Chen2022}. A lack of a radius gap (i.e., the observed unimodal planet radius distribution) would suggest that the ``typical'' population of sub-Neptunes, comprised of systems with rocky cores and H/He atmospheres, may be absent from the $\leq$100 au sample. This suggests either that the larger planets in the $\leq$100 au sample could be unusually large rocky planets that have stripped atmospheres, or that they are another population such as water worlds, which could produce a smooth transition between the super-Earth and sub-Neptune regimes. Radial velocity measurements to constrain the masses (and thus the bulk densities) of the larger planets (those with $R_{p} > 2$ \rearth) in the 100 au sample will be crucial to understand the composition of these planets and explain the unimodal planet radius distribution of the $\leq$100 au sample. However, after determining which star each planet in the sample orbits, a radius gap may become apparent.

\section{Summary and Future Work}
We have assembled a sample of 283 transiting exoplanets identified by the Kepler mission, hosted by 207 binary systems identified in community follow-up \citep{Furlan2017} and our own AO imaging using Keck/NIRC2. We observed the systems with LRS2 on the HET to obtain spectroscopy, then revised the stellar and planetary properties while accounting for the observational biases caused by the presence of a second star in the planetary system. We investigated the radius distribution in these small planets and found that the planet radius distribution is dependent on the binary star separation. The population of planets around the closest binaries (those with separations $\rho \leq 100$ au) have a suppressed population of sub-Neptunes, with only a single apparent peak in the radius distribution near the super-Earth occurrence peak of 1.3 \rearth. 

Our observational results support a theoretical expectation that protoplanetary disk mass and lifetime directly impact the small planet radius distribution by inhibiting sub-Neptune formation. However, several questions remain. Two-thirds of binaries may never form planets \citep{Kraus2016, Moe2021}, while the one-third that do form planets may have normal disk lifetimes \citep[e.g.,][]{Kraus2012}, motivating studies establishing which binary properties influence planet survival. The composition of the planets in the long sub-Neptune tail of the close binary radius distribution should be explored to establish whether they are water worlds, large rocky cores, or gaseous Neptune-like planets. Finally, larger samples of binaries will reveal whether the location of the radius gap is constant or shifts to smaller planetary radii in closer binaries.

Crucial inputs to further studies of planets in binary systems that can address these questions will include assembling larger samples of planets in binary systems (via missions like PLATO \citep{Rauer2014}, the Nancy Grace Roman Space Telescope \citep{Akeson2019}, or Earth 2.0 \citep{Ge2022}, plus follow-up for multiplicity using Gaia and high-resolution imaging) for more robust demographics; additional software development to enable more robust and reliable derivation of system properties for planet-host binary systems, including radial velocity measurements; and calculating occurrence rates to understand intrinsic properties of the population of planets in binary star systems. However, the results of this work demonstrate that analysis of planet-hosting binary stars can begin to provide insights into planet formation even without occurrence rate calculations.

\facilities{Keck:II (NIRC2), HET (LRS2), Exoplanet Archive}
\software{astropy \citep{astropy2013, astropy2018, astropy2022}, corner \citep{Foreman-Mackey2016}, dustmaps \citep{Green2018}, emcee \citep{Foreman-Mackey2013}, matplotlib \citep{Hunter2007}, numpy \citep{Harris2020}, pyphot \citep{pyphot}, scipy \citep{Virtanen2020}, scikit-learn \citep{scikit-learn}}

We thank the referee for their comments, which have improved this manuscript. We thank Natalie Batalha and Anne Dattilo for useful discussions regarding the results of this work, and we thank Daniel Krolikowski for generating the \texttt{TelFit} models for our telluric correction.

The authors sincerely thank the observing staff and resident astronomers at the Hobby-Eberly Telescope for obtaining some of the observations presented in this work. We acknowledge the Texas Advanced Computing Center (TACC) at The University of Texas at Austin for providing high performance computing, visualization, and storage resources that have contributed to the results reported within this paper.

EG was supported by NASA Grants 80NSSC20K0957 (Exoplanets Research Program) and 80NSSC22K0295 (\emph{TESS} Guest Observer Cycle 4).  
TJD acknowledges support from UKRI STFC AGP grant ST/W001209/1. DH acknowledges support from the Alfred P. Sloan Foundation, the National Aeronautics and Space Administration (80NSSC22K0781), and the Australian Research Council (FT200100871).
ALK was supported by the NASA Exoplanets (XRP) grant 80NSSC22K0781.
For the purpose of open access, the author has applied a Creative Commons Attribution (CC BY) license to any Author Accepted Manuscript version arising from this submission.

The Hobby-Eberly Telescope (HET) is a joint project of the University of Texas at Austin, the Pennsylvania State University, Ludwig-Maximilians-Universität München, and Georg-August-Universität Göttingen. The HET is named in honor of its principal benefactors, William P. Hobby and Robert E. Eberly. The Low-Resolution Spectrograph 2 (LRS2) was developed and funded by the University of Texas at Austin McDonald Observatory and Department of Astronomy and by Pennsylvania State University. We thank the Leibniz-Institut für Astrophysik Potsdam (AIP) and the Institut für Astrophysik Göttingen (IAG) for their contributions to the construction of the integral field units. 

Some of the data presented herein were obtained at the W. M. Keck Observatory, which is operated as a scientific partnership among the California Institute of Technology, the University of California and the National Aeronautics and Space Administration. The Observatory was made possible by the generous financial support of the W. M. Keck Foundation. The authors wish to recognize and acknowledge the very significant cultural role and reverence that the summit of Maunakea has always had within the indigenous Hawaiian community.  We are most fortunate to have the opportunity to conduct observations from this mountain. This research has made use of the Keck Observatory Archive (KOA), which is operated by the W. M. Keck Observatory and the NASA Exoplanet Science Institute (NExScI), under contract with the National Aeronautics and Space Administration.

This publication makes use of data products from the Two Micron All Sky Survey, which is a joint project of the University of Massachusetts and the Infrared Processing and Analysis Center/California Institute of Technology, funded by the National Aeronautics and Space Administration and the National Science Foundation. This research has made use of the SVO Filter Profile Service (\url{http://svo2.cab.inta-csic.es/theory/fps/}) supported from the Spanish MINECO through grant AYA2017-84089. This research has made use of the VizieR catalogue access tool, CDS, Strasbourg, France (DOI : 10.26093/cds/vizier). The original description of the VizieR service was published in 2000, A\&AS 143, 23. This work has made use of data from the European Space Agency (ESA) mission {\it Gaia} (\url{https://www.cosmos.esa.int/gaia}), processed by the {\it Gaia} Data Processing and Analysis Consortium (DPAC, \url{https://www.cosmos.esa.int/web/gaia/dpac/consortium}). Funding for the DPAC has been provided by national institutions, in particular the institutions participating in the {\it Gaia} Multilateral Agreement. This research has made use of the Exoplanet Follow-up Observation Program website, which is operated by the California Institute of Technology, under contract with the National Aeronautics and Space Administration under the Exoplanet Exploration Program. This research has made use of the NASA Exoplanet Archive, which is operated by the California Institute of Technology, under contract with the National Aeronautics and Space Administration under the Exoplanet Exploration Program.

\begin{longrotatetable} %
\begin{deluxetable*}{ccCCCCCCCCCCCCCC}
\setcounter{table}{1}
\tablecaption{System Parameters for Each Source \label{tab:obs}}
\tablecolumns{14}
\tablehead{
\colhead{KOI} & \colhead{$\rho$} & \colhead{Obs date} & \colhead{r'} & \colhead{S/N} & \colhead{$\Delta m_{i}$} & \colhead{$\Delta m_{LP600}$} & \colhead{$\Delta m_{Gaia}$} & \colhead{$\Delta m_{562 nm}$} 
& \colhead{$\Delta m_{692 nm}$} & \colhead{$\Delta m_{880 nm}$} & \colhead{$\Delta m_{J}$} & \colhead{$\Delta m_{K}$} & \colhead{References}\\
\colhead{} & \colhead{(\arcsec)} & \colhead{} & \colhead{(mag)} & \colhead{} & \colhead{(mag)} & \colhead{(mag)} & \colhead{(mag)} & \colhead{(mag)} 
& \colhead{(mag)} & \colhead{(mag)} & \colhead{(mag)} & \colhead{(mag)} & \colhead{}\\
}
\startdata  
0042 & 1.66 & 20220713 & 9.33 & 2095 & \nodata & $3.04 \pm 0.17$ & \nodata & $4.24 \pm 0.15$ & \nodata & \nodata & $2.21 \pm 0.03$ & $1.87 \pm 0.02$ & F17 A12 H11 B16 K16 \\
0112 & 0.11 & 20230924 & 12.74 & 521 & \nodata & \nodata & \nodata & \nodata & $1.78 \pm 0.15$ & $1.63 \pm 0.15$ & $0.86 \pm 0.03$ & $1.02 \pm 0.23$ & F17 A12 H11 K16 \\
0162 & 0.29 & 20221021 & 13.77 & 297 & \nodata & $0.81 \pm 0.29$ & \nodata & \nodata & \nodata & \nodata & \nodata & \nodata & F17 \\
0163 & 1.22 & 20230910 & 13.49 & 324 & \nodata & $0.36 \pm 0.03$ & $0.43 \pm 0.05$ & \nodata & \nodata & \nodata & \nodata & $0.17 \pm 0.00$ & F17 \\
0165 & 0.28 & 20230925 & 13.90 & 257 & \nodata & \nodata & \nodata & \nodata & \nodata & \nodata & \nodata & $0.86 \pm 0.00$ & F17 \\
\enddata
\tablecomments{The full table is available in machine-readable format online. The table columns include the KOI number, the angular separation, the observation date, the system r' magnitude, the S/N of the observation, and any observed contrasts, including literature references if relevant. F17 is \citet{Furlan2017}; A12 is \citet{Adams2012}; H11 is \citet{Howell2011}; B16 is \citet{Baranec2016}; LB12 is \citet{Lillo-Box2012}; LB14 is \citet{Lillo-Box2014}; and K16 is \citet{Kraus2016}. Any unmarked K magnitudes are from this work.}
\end{deluxetable*}
\end{longrotatetable}

\include{obstable_nirc2_edit_rgap2_aastex}

\bibliography{mcmc4_bib}
\end{document}